\documentclass[10pt,english]{article}
\usepackage[paperheight=9in,paperwidth=6in,top=0.6in,bottom=0.6in,right=0.7in,left=0.7in]{geometry}

\usepackage[T1]{fontenc}
\usepackage[latin9]{inputenc}
\usepackage{amsmath,amsfonts,amsthm}
\usepackage{amssymb}
\usepackage{graphicx}
\usepackage{esint}
\usepackage{tikz}
\usepackage{physics}
\usepackage{hyperref}
\makeatletter

\usepackage{babel}

\begin{document}
{\renewcommand{\thefootnote}{\fnsymbol{footnote}}
\medskip

\begin{center}
{\LARGE Quantum Walk on a Spin Network}
\bigskip

\vspace{1.5em}
Klee Irwin$^{a}$\footnote{email: Klee@QuantumGravityResearch.org},
Marcelo M. Amaral$^{ab}$\footnote{email: mramaciel@gmail.com},
Raymond Aschheim$^{a}$\footnote{email: Raymond@QuantumGravityResearch.org},

\vspace{0.5em}
\small \em $^a$Quantum Gravity Research\\
\small \em Los Angeles, CA\\
\small \em $^b$Institute for Gravitation and the Cosmos, \\
\small \em Pennsylvania State University, University Park, PA 
\vspace{1.5em}
\end{center}
}

\setcounter{footnote}{0}

\begin{abstract}
\normalsize
We apply a discrete quantum walk from a quantum particle on a discrete quantum spacetime from loop quantum gravity and show that the related entanglement entropy drives an entropic force. We apply these concepts in a model where walker positions are topologically encoded on a spin network.

\end{abstract}

\section{Introduction}

One of the principal results from Loop Quantum Gravity (LQG) is a discrete spacetime$-$a network of loops implemented by spin networks \cite{Rovelli-Vidotto-Book} acting as the digital/computational substrate of reality. 
In order to better understand this substrate, it is natural to use tools from quantum information / quantum computation.
Gravity, from a general perspective, has been studied with thermodynamic methods. In recent years, numerous questions on black hole entropy and entanglement entropy 
have made this an active field of research. 
In terms of quantum information and quantum computation, advances have been achieved with the aid of many new mathematical tools. Herein, we present the development of one such tool, which we call the discrete-time quantum walk (DQW).
We will see that the problem of a quantum particle on a fixed spin network background from LQG
can be worked out with the DQW. This gives rise to a new understanding of entanglement entropy and entropic force,
permitting the proposal of a model for dynamics. 
In terms of physical ontology, we suggest dynamics and mass emerge from this spin network topology, as implemented by the DQW.

In summary, we will reinterpret some results from LQG that emphasizes the quantum information perspective of a quantum-geometric spacetime. That is, it adopts Wheeler's \textit{it from bit} and the newer \textit{it from qubit} ontologies$ - $the general digital physics viewpoint.

\section{Methods}

\subsection{Particle interacting LQG}
\label{particleinteractingLQG}
We start by considering a quantum particle on a quantum gravitational
field from LQG \cite{Rovelli-Vidotto}. In LQG, spin networks define
quantum states of the gravitational field. To consider a quantum particle
on this gravitational field, we consider the state space built from
tensor product of the gravity state $\mathcal{H}_{LQG}$ and the particle
state $\mathcal{H}_{P}$, $\mathcal{H}=\mathcal{H}_{LQG}\otimes\mathcal{H}_{P}$. The
LQG state space can be spanned by a spin network basis $\ket{s}$ that
is a spin network graph $\Gamma=(V(\Gamma),L(\Gamma))$, with $V(\Gamma)$ 
vertices coloring which elements denoted by $v_{1},v_{2}...,$ and $L(\Gamma)\rightarrow l_{1},l_{2}...,$ ($\{\frac{1}{2},1,\frac{3}{2}...\}$)
links coloring. For the particle state space the relevant contribution
comes from the position on the vertices of graph $\Gamma$, spanned
by $\ket{x_{n}}$ where ($n=1,2...$). The quantum state of the particle captures
information from discrete geometry and cannot be considered independently
from it. Therefore the Hilbert space is spanned by $\ket{s,x_{n}}$ and the
Hamiltonian operator can be derived by fixing a spin network to the graph, and calculating
the matrix element of this operator $\expval{H}{\psi}$ with $\psi\in H_{P}$.
Accordingly, for the interaction between the particle and the fixed
gravity state space \cite{Rovelli-Vidotto} we have,
\begin{equation}
 \expval{H}{\psi}=\kappa\underset{l}{\sum}j_{l}(j_{l}+1)\left(\psi(l_{f})-\psi(l_{i})\right)^{2},\label{eq:RovelliHamiltonian}
\end{equation}
where $\kappa$ is a constant that we can initially take as equal to
one, $l_{f}$ are the final points of the link $l$ and $l_{i}$ the
initial points. The interaction term takes this form because the relevant Hilbert space depends on the wave functions at the vertices.
So if the link $l$ starts at vertex $m$ and ends at vertex
$n$ we can change the notation, relabeling the color of this link $l$
between $m$ and $n$ as $j_{mn}$ and the wave function on the end points as 
$\psi(v_{m})$, $\psi(v_{n})$. Now, for $H$, we have
\begin{equation}
\expval{H}{\psi}=\kappa\underset{l}{\sum}j_{mn}(j_{mn}+1)\left(\psi(v_{n})-\psi(v_{m})\right)^{2},\label{eq:IsaiasHamiltonian}
\end{equation}

The operator $H$ is positive semi-definite. The ground state
of $H$ corresponds to the case where the particle is maximally delocalized.
This leads to entropy \cite{Rovelli-Vidotto}. In reference \cite{Garcia-Islas}
it was considered that the classical random walk is associated with (\ref{eq:IsaiasHamiltonian}), a Markov chain. The transition probabilities for this random walk are
\begin{equation}
P_{mn}=\frac{j_{mn}(j_{mn}+1)}{\underset{k}{\sum}j_{mk}(j_{mk}+1)}.\label{eq:IsaiasProbabilitie}
\end{equation}
and \cite{Garcia-Islas} shows that this reproduces an entropic force. This random walk is implemented with
the Laplacian solution (\ref{eq:IsaiasHamiltonian}) and differs from the Laplacians coming from the
discrete calculus work out in \cite{ThurigenLaplacian}. We will consider DQW relations with
more general Laplacians in future work.

\subsection{DQW}

As study the influence of discrete geometry on a quantum
field. Here, it is natural to consider the quantum version of random
walks. To have DQWs we need an auxiliary subspace, the coin toss space,
to define the unitary evolution 
\begin{equation}
U=S(C\otimes I),\label{eq:Uquantumwalkregulargraph}
\end{equation}
where $S$ is a swap operation that changes the position to a neighbor
node and $C$ is the coin toss operator related to this auxiliary subspace.
On some graphs its clear how to define the coin toss Hilbert space 
\cite{Aharonov-Ambainis} but for the spin network
considered above its not clear exactly what auxiliary space to use.
For a more general approach, we can make use of Szegedy's DQW
\cite{Szegedy1, Szegedy2}, which we can utilize in two ways: 1- consider a bipartite
walk or 2 - a walk with memory. For a bipartite walk, if we consider
a graph $\Gamma$, we can simply make an operation of duplication to
obtain a second graph $\bar{\Gamma}$. For our purpose here, it is
better to consider the second option $-$ the walk with memory, considering
the $N_{v}$-dimensional Hilbert space $\mathcal{H}_{n}$, $\{\ket{n},n=1,2,...,N_{v}\}$ 
and $\mathcal{H}_{m}$, $\{\ket{m},m=1,2,...,N_{v}\}$, 
where $N_{v}$ is the number of the vertex $V(\Gamma)$ . The state
of the walk is given as the product $\mathcal{H}_{n}^{N_{v}}\otimes \mathcal{H}_{m}^{N_{v}}$ spanned
by these bases. That is, by states at the previous $\ket{m}$ and
current $\ket{n}$ steps, defined by 
\begin{equation}
\ket{\psi_{n}(t)}=\underset{m}{\overset{N_{v}}{\sum}}\sqrt{P_{mn}}\ket{n}\otimes\ket{m},\label{eq:SzegedyState}
\end{equation}
where $P_{mn}$ is the transition probabilities that define a classical
random walk, a Markov chain, which is a discrete time stochastic process
without a memory, with 
\begin{equation}
\underset{n}{\sum}P_{mn}=1.\label{eq:TransitionPequal1}
\end{equation}
Note that (\ref{eq:TransitionPequal1}) is implied by definition of (\ref{eq:IsaiasProbabilitie}).
For the evolution, we can consider a simplified version of Szegedy's
DQW \cite{Bkollartesi} by defining a reflection, which we can interpret with the
unitary coin toss operator\footnote{Szegedy's DQW does not generally use a coin toss operator in the literature}, 
\begin{equation}
C=2\underset{n}{\sum}\ket{\psi_{n}}\bra{\psi_{n}}-I,\label{eq:SzegedyCoin}
\end{equation}
and a reflection with inverse action of the $P$ swap (previous and current
step) that can be implemented by a generalized swap operation
\begin{equation}
S=\underset{n,m}{\sum}\ket{m,n}\bra{n,m},\label{eq:SzegedySwap}
\end{equation}
where we have the unitary evolution 
\begin{equation}
U=CS,\label{eq:SzegedyEvolution}
\end{equation}
that defines the DQW.

So using Szegedy's approach is a straightforward way to obtain the discrete
quantum walk on the spin network $\Gamma$ considered above. It is
given by equations (\ref{eq:SzegedyState} - \ref{eq:SzegedyEvolution})
with $P$ given by (\ref{eq:IsaiasProbabilitie}). Namely, for equation (\ref{eq:SzegedyState})
\begin{equation}
\ket{\psi_{n}(t)}=\underset{m}{\overset{N_{v}}{\sum}}\sqrt{\frac{j_{mn}(j_{mn}+1)}{\underset{k}{\sum}j_{mk}(j_{mk}+1)}}\ket{n}\otimes\ket{m}.\label{eq:SzegedyState-1}
\end{equation}
We can interpret the coin toss space as the space of decisions encapsulating
a nondeterministic process possessing memory, such that we have a unitary
evolution. This approach makes the usual entropy convert into entanglement
entropy between steps in time. Szegedy's DQW is a general algorithm
quantizing a Markov chain defined by transition probabilities $P_{n,m}$.
These transition probabilities are obtained directly from the Hamiltonian
of the system such that Szegedy's DQW is used to simulate 
this system with quantum computation \cite{child2010}.

\subsection{Entanglement Entropy and Entropic Force}

We turn now to calculate entanglement entropy. Consider the Schmidt
decomposition. Take a Hilbert space $\mathcal{H}$ and decompose it into
two subspaces $\mathcal{H}_{1}$ of dimension $N_{1}$ and $\mathcal{H}_{2}$
of dimension $N_{2}\geq N_{1}$, so 
\begin{equation}
\mathcal{H}=\mathcal{H}_{1}\otimes\mathcal{H}_{2}.\label{eq:HilbertspaceDecom}
\end{equation}
Let $\ket{\psi}\in\mathcal{H}_{1}\otimes\mathcal{H}_{2}$, and $\{\ket{\psi_{i}^{1}}\}\subset\mathcal{H}_{1}$,
$\{\ket{\psi_{i}^{2}}\}\subset\mathcal{H}_{2}$, and positive real numbers
$\{\lambda_{i}\}$, then the Schmidt decomposition would be 
\begin{equation}
\ket{\psi}=\underset{i}{\overset{N_{1}}{\sum}}\sqrt{\lambda_{i}}\ket{\psi_{i}^{1}}\otimes\ket{\psi_{i}^{2}},\label{eq:Schmidtdecomp}
\end{equation}
where $\sqrt{\lambda_{i}}$ are the Schmidt coefficients and the number of
the terms in the sum is the Schmidt rank, which we label $N$.
With this, we can calculate the entanglement entropy between the two subspaces
\begin{equation}
S_{E}(1)=S_{E}(2)=-\underset{i\in N}{\sum}\lambda_{i}log\lambda_{i}.\label{eq:defEntanglementEntropy}
\end{equation}
We can now calculate the local entanglement entropy between the previous $m$
step and the current $n$ step (similar for current and next steps). Identifying
the Schmidt coefficients $\sqrt{\lambda_{i}}$ with $\sqrt{P_{mn}}$, with $P_{mn}$ given by (\ref{eq:IsaiasProbabilitie})
and, from (\ref{eq:SzegedyState-1}), we see that the Schmidt rank $N$
is the valence rank of the node. Then insert in (\ref{eq:Schmidtdecomp},
\ref{eq:defEntanglementEntropy}), and the local entanglement entropy
on current step is 
\begin{equation}
S_{E_{n}}=-\underset{m}{\overset{N}{\sum}}P_{mn}logP_{mn}.\label{eq:EntanglementEntropyQWtime}
\end{equation}
By maximizing Entanglement Entropy 
\begin{equation}
S_{E_{n}}=logN_{max},\label{eq:EntanglementEntropyQWtimeMax}
\end{equation}
where $N_{max}$ is the largest valence. Which gives the entropic force
for gravity worked out in \cite{Garcia-Islas,Verlinde} with $\frac{dS}{dx}=|S_{E_{n}}-S_{E_{m}}|$
proportional to a small number identified with the particle mass
$M$ 
\begin{equation}
\frac{dS}{dx}=|S_{E_{n}}-S_{E_{m}}|=\alpha M,\label{eq:deltaEntropyequalMass}
\end{equation}
where, if we take the logarithm of (\ref{eq:EntanglementEntropyQWtimeMax})
to be base 2, $\alpha$ is a constant of dimension $[bit/mass]$.

Each of these interpretations has related applications. For example, with the DQW, we
have unitary evolution by encoding a non-deterministic part of the
classical Markov chain. This gives an internal structure for the particle as well as the 
entanglement entropy value. In such a digital physics substrate, the particle
walks in such a way as to maximize its entanglement entropy based upon the inherent memory of
its walking path (a measurement of the system on $n$ implies the state at previous $m$), which generates an entropic force.

Interestingly, we can consider von Neumann Entropy (\ref{eq:EntanglementEntropyQWtime})
in the context of quantum information. The probabilities are the Markov
chain connecting the steps. Accordingly, the particles construct ``letters'' of a spatiotemporal code
as expression of the allowed restricted but non-determined walking paths, where the entropy measures the information needed. From this, we see how the entropic force emerges.

\section{Results}

\subsection{Entropy of Black Hole}
\label{EntropyofBlackHole}
The framework above is a discretized quantum field expressing phases as algebraically allowed patterns upon fixed discrete geometry. Specifically a moduli space algebraic stack. Gravitational entropic force suggests a unified picture of gravity and matter via a quantum gravity approach. Consider now a regime from pure quantum gravity, like the
black hole quantum horizon, so that there are no quantum fields but only
quantum geometry. We can use our DQW to simulate this regime.

Again, DWQs encode local entanglement entropy (\ref{eq:EntanglementEntropyQWtime})
in the sense that the chains or network of step paths are always dynamically under construction as a "voxelated" animation. This resembles the isolated quantum horizons
formulation of LQG \cite{Rovelli:1996dv,Ashtekar:2000eq,Domagala:2004jt,Ghosh:2004wq,G.:2015sda}, which gives explains the origin of black hole entropy in LQG. In
this scenario, the Horizon area emerges from the stepwise
animated actions at the Planck scale that are simulated by
DQWs. We argue that DQWs are the Planck scale substrate forming the emergent black hole quantum horizon, where particle masses composite to black hole mass in the full aggregate of quantum walks on the spin network of the event horizon.

In the isolated quantum horizons formulation, entropy is generally 
calculated by considering the eigenvalues of the area operator $A(j)$
and introducing an area interval $\delta a=[A(j)-\delta,A(j)+\delta]$ 
of the order of the Planck length with a relation to the classical area $a$ of the horizon. $A(j)$ is given
by 
\begin{equation}
A(j)=8\pi\gamma l_{p}^{2}\underset{l}{\sum}\sqrt{j_{l}(j_{l}+1)},
\end{equation}
where $\gamma$ is the Barbero-Immirzi parameter and $l_{p}$ the
Planck length. The entropy, in a dimensional form, is 
\begin{equation}
S_{BH}=lnN(A),\label{eq:blackholeentropyklnN}
\end{equation}
with $N(A)$ the number of micro-states of quantum geometry on the
horizon (area interval $a$) implemented, combinatorially, by considering states with
link sequences that implement the two conditions 
\begin{equation}
8\pi\gamma l_{p}^{2}\overset{N_{a}}{\underset{l=1}{\sum}}\sqrt{j_{l}(j_{l}+1)}\leq a,\label{eq:conditionBHentropy1}
\end{equation}
related to the area, where $N_{a}$ is the number of admissible $j$ that puncture the
horizon area and 
\begin{equation}
\overset{N_{a}}{\underset{l=1}{\sum}}m_{l}=0\label{eq:conditionBHentropy2}
\end{equation}
related to the flux with $m_{l}$, the magnetic quantum number satisfying
the condition $-j_{l} \leq m_{l} \leq j_{l}$.

The detailed calculation \cite{Ashtekar:2000eq,G.:2015sda} shows
that the dominant contribution to entropy comes from states in which
there is a very large number of punctures. Thus, it is productive to
interpret this entropy as quantum informational entropy (\ref{eq:EntanglementEntropyQWtime}).
Let us investigate how the horizon area and related entropy emerges
from maximal entanglement entropy of DQWs. Condition (\ref{eq:conditionBHentropy1})
is associated with these DQWs.
From (\ref{eq:EntanglementEntropyQWtime}
and \ref{eq:EntanglementEntropyQWtimeMax}), considering edge coloring,
maximal entanglement entropy occurs for states on nodes of large
valence $N_{max}$ and sequence with $j_{l}=\frac{l}{2}$ ($l=1,2,...,N_{max})$. Accordingly, for the entropy calculation, it is admissible that $j$, which punctures the
horizon area, respects these sequences and that associated nodes have large valence rank.
Therefore, we can rewrite condition (\ref{eq:conditionBHentropy1}) as
\begin{equation}
\overset{N_{a_{i}}}{\underset{i=1}{\sum}}a_{i}=a_{c},\label{eq:conditionBHentropyA}
\end{equation}
where $N_{a_{i}}$ is the number of admissible nodes with 
\begin{equation}
a_{c}=\frac{a}{4\pi\gamma l_{p}^{2}}.\label{eq:conditionBHentropy1-1-1}
\end{equation}
and each $a_{i}$ is calculated from 
\begin{equation}
a_{i}=\overset{N_{max}}{\underset{l=1}{\sum}}\sqrt{l(l+2)},\label{eq:conditionBHentropy1-1}
\end{equation}
and considering the dominant contributions given by the over-estimate
each $a_{i}$ and counting $N(a_{c})$ that will give $N(A)$, each $a_{i}$ is an integer
strictly greater than $1$ 
\begin{equation}
\sqrt{l(l+2)}=\sqrt{(l+1)^{2}-1}\approx l+1,\label{eq:conditionBHentropy1-1-2}
\end{equation}
which means that the combinatorial problem we need to solve is to
find $N(a_{c})$ such that (\ref{eq:conditionBHentropyA}) holds.
This was discussed in a similar problem in \cite{Rovelli:1996dv}. It is  
straightforward\footnote{Because the cardinal $N(a_{c})$ of the set of ordered tuples of integers strictly greater than $1$ summing to $a_{c}$ is the $a_{c}^{th}$ Fibonacci number $F(a_{c})$.} 
to see that 
\begin{equation}
logN(A)=\frac{log(\phi)}{\pi\gamma}\frac{a}{4l_{p}^{2}},\label{eq:blackholeentropylogNphi}
\end{equation}
where $\phi=\frac{1+\sqrt{5}}{2}$ is the golden ratio%
\footnote{The golden ratio is an irrational number (1.61803398875...) as the solution
to the equation $\phi^{2}=\phi+1$.%
}.

With this established, we can now conjecture that this entropy, relating to states that give maximal
entanglement entropy, is simply the entanglement entropy of the DQWs. Accordingly, holographic quantum horizons can be simulated by the DQWs with maximal internal
entanglement entropy. The walker moves are from node $m$ to node $n$, each each with large valence number so that equation (\ref{eq:deltaEntropyequalMass})
works. If we consider a walker mass spanning $I$ with Planck mass $m_{p}$,
we can choose the proportionality constant so that
\begin{equation}
\triangle S_{E}=|S_{E_{n}}-S_{E_{m}}|=I,\label{eq:deltaEntropyequalMass-1}
\end{equation}
so $M=Im_{p}$ gives the amount of information on the horizon, and $\triangle S_{E}$
is given in $bits$. We can make explicit the $log_{2}$ on (\ref{eq:blackholeentropylogNphi})
so that it is given in $bits$ too and propose that $log_{2}N(A)=\triangle S_{E}$
and that this measure of information gives the emergent Bekenstein-Hawking
entropy. DQWs encode or express black hole information, so to infer its
entropy, one needs to make contact with some frame of DQWs. The
particles that simulate black holes will be more probable on a node
of maximal entanglement entropy encapsulating $logN(A)$. Note that
$logN(A)$ is not the usual statistical entropy because we have
not taken into account all of the micro-states or the
flux condition. From (\ref{eq:blackholeentropyklnN})
the black hole entropy, changing the logarithm, is 
\begin{equation}
S_{BH}=log_{2}N(A)=\frac{log_{2}(\phi)}{\pi\gamma}\frac{a}{4l_{p}^{2}}.\label{eq:blackholeentropyklnN-1}
\end{equation}
Therefore, Bekenstein-Hawking entropy is recovered by setting the Barbero-Immirzi
parameter 
\begin{equation}
\gamma=\frac{log_{2}(\phi)}{\pi},\label{eq:gammaphipi}
\end{equation}
showing that the states of maximal entanglement entropy are dominant
in black hole entropy. So we can think of Bekenstein-Hawking entropy
as emergent from the local entanglement entropy above. A horizon area
$a$ spanning $4I$ Planck areas has $I$ $bits$ like (\ref{eq:deltaEntropyequalMass-1}).

\subsection{A model of walker position topologically encoded on a spin network}

\begin{figure}[!ht]
	\centering{}
	\includegraphics[scale=0.50]{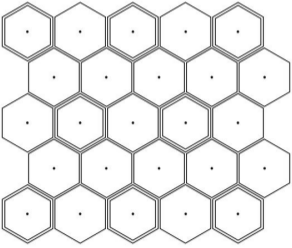}
 \caption{Loops.} \label{qwspinfoamimg01}
\end{figure}

The Clebsh-Gordan condition at each node is realized by
covering the graph with loops.
From (\ref{eq:IsaiasProbabilitie}) and (\ref{eq:EntanglementEntropyQWtime}),
we can compute the local entropy from a vertex as
\begin{equation}
S_{E_{n}}=log\sigma-\frac{1}{\sigma}\underset{m}{\overset{N}{\sum}}j_{mn}(j_{mn}+1)log\left(j_{mn}(j_{mn}+1)\right),\label{eq:EntanglementEntropyQWlocalformtoymodel}
\end{equation}
where $\sigma=\underset{m}{\overset{N}{\sum}}j_{mn}(j_{mn}+1)$ of
neighbor links. For example, at a node $\{2,3,3\}$, $j=\{1,\frac{3}{2},\frac{3}{2}\}$
gives $\sigma=\frac{19}{2}$ and $S_{E_{n}}=1.06187$. At a node $\{2,2,2\}$,
$j=\{1,1,1\}$ gives $\sigma=6$, $S_{E_{n}}=1.09861$, which is the
maximum possible local entropy. Note that (\ref{eq:EntanglementEntropyQWlocalformtoymodel})
is the local entropy formula given in \cite{Garcia-Islas} as 14 and is in accordance
with well known LQG formulas for quantized length and area. See figure (\ref{qwspinfoamimg01}).

The local entropy at each node is color coded. From equation (\ref{eq:deltaEntropyequalMass}), 
a massless particle moves on the same color and a massive particle moves along
constant absolute color differences.

\begin{figure}[!ht]
	\centering{}
	\includegraphics[scale=0.50]{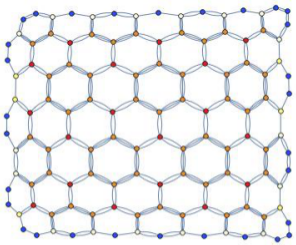}
 \caption{Entropy is color coded.} \label{qwspinfoamimg02}
\end{figure}

In the figure (\ref{qwspinfoamimg02}) :\protect \\
(blue)$\{l_{1},l_{2},l_{3}\}=\{0,1,1\}$ or $\{0,2,2\}$ (side effect)\protect \\
(white) ... $\{1,2,3\}$\protect \\
(yellow) ... $\{1,1,2\}$\protect \\
(orange) ... $\{2,3,3\}$\protect \\
(red) ... $\{2,2,2\}$.\protect \\

Dynamics:\protect \\
Photons can orbit on orange hexagons.
No possible particles with constant mass. 
Possible travel of massive particle, with 
$mass=|S_{(red)}-S_{(orange)}|$, ($=0.037$) 
interacting with photons at some orange vertex.

\begin{figure}[!ht]
	\centering{}
	\includegraphics[scale=0.50]{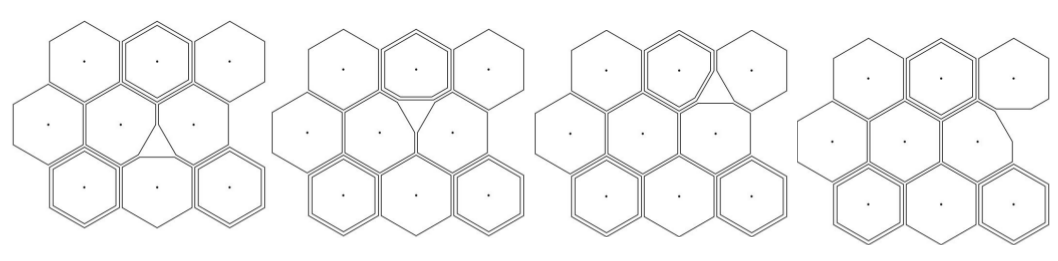}
 \caption{Particle and Pachner moves.} \label{qwspinfoamimg03}
\end{figure}

The walker position or the presence of a particle at one node is
encoded by a triangle. Its move is a couple of 3-1 and 1-3
Pachner moves on neighbor positions, piloted by the walk
probability. See figure (\ref{qwspinfoamimg03}).

 \section{Conclusion}
  We have presented a compelling idea that
we can apply the results and tools from quantum information and quantum
computation to a quantum spacetime code theoretic view using algebraic graph formalism. We considered a DQW of a quantum particle on a quantum
gravitational field and studied applications of related entanglement 
entropy. This memory time based entanglement entropy drives an
entropic force, suggesting a unified picture of gravity and matter. Following this,
we proposed a model for walker positions topologically encoded on a
spin network, which can easily be re-expressed using twistors. This results in anomaly cancellation because the particles
are no longer points but Planck scale voxels, as tetrahedral units of spacetime. 
We note that more
complex models can be build, with more dimension and 8 quantum numbers
in E8 model \cite{Aschheim} and models with realistic emergent masses
can be explored by computation.

A better understanding of Entanglement Entropy on black hole and application
on cosmology are under investigation, as well as relating this model
with QuasiCrystal E8 model \cite{Fang-Irwin}: the spin network can
be choosen as the dual of a quasicrystal and the digital physics rules
can be implemented by the quantum walk.

\end{document}